\documentclass[preprint,12pt]{article}
\usepackage{graphics}
\usepackage{authblk}
\usepackage{epstopdf}
\usepackage{amsmath}
\usepackage{amssymb}
\usepackage[round]{natbib}
\usepackage{graphicx}
\usepackage{caption}
\usepackage{subcaption}
\usepackage{color}
\usepackage{graphicx}
\begin{document}
\bibliographystyle{plainnat}
\title{Ambipolar decay of magnetic field in magnetars and the observed magnetar
activities}

\author[1]{Badal Bhalla}
\author[1]{Monika Sinha}
\affil[1]{Indian Institute of Technology Jodhpur, India}

\maketitle
\begin{abstract}
Magnetars are comparatively young neutron stars with ultra-strong 
surface magnetic field in the range $10^{14}-10^{16}$ G. The old neutron
stars have surface magnetic field some what less $\sim 10^8$ G 
 which clearly indicates the decay of field with time. One possible way of 
magnetic field decay is by ambipolar diffusion. We describe the
general procedure to solve for the ambipolar velocity inside the star core without any approximation. With a realistic model of neutron 
star we determine the ambipolar velocity configuration inside 
the neutron star core and hence find the ambipolar decay rate, associated time scales and the magnetic energy dissipated  which is consistent with the magnetar observations.
\end{abstract}

%
% Uncomment for keywords
\vspace{2pc}

\section{Introduction}

The unique astrophysical objects neutron stars are mostly observed 
as radio pulsars. Though from neutron stars we receive radiation 
in all ranges of the electromagnetic spectrum, radiation in radio, X-ray 
and sometime $\gamma$-ray are dominating. Neutron stars are 
magnetized. Most of the radio pulsars are isolated neutron stars 
with rotation periods 100 ms to 10 s and typical surface magnetic 
field of the order $10^{10}-10^{13}$ G. Few of them are in binary 
systems. Another group of pulsars have shorter periods and 
the pulsars with rotation periods less than $100$ ms are known 
as milli-second pulsars. Most of them have typical surface magnetic
field range $10^{8}-10^{10}$ G and are present in binary system. They are
more aged than isolated radio pulsars. Another special kind of pulsars
are anomalous X-ray pulsars. They are called ``anomalous" because
like ordinary pulsars they are neither rotation powered nor in
binary system to be accretion powered. They are generally believed
to have very strong surface magnetic field in the range of $10^{14}-10^{16}$ G 
and their activity is due to magnetic energy dissipation. 
Another class of objects, the soft gamma repeaters, are also
believed to come under the same category of this type of neutron
stars with ultra-strong surface magnetic field. This class 
of objects are known as magnetars and are younger compared to
other class of neutron stars. %The observation of this nice 
%connection of surface magnetic field strength with age of the
%neutron stars shows the signature of neutron star magnetic field
%evolution. Along with this, 
The emission from magnetars is %also
believed to be caused by the magnetic field evolution inside 
magnetars. In %all these 
this respect the internal magnetic
field evolution is very important phenomenon to study.

The magnetic field evolution can occur through several processes: (a) Ohmic decay, (b) Ambipolar diffusion and (c) Hall-drift.
The magnetar activity is generally explained by the field evolution
inside the crust of the neutron star. In the core, the Ohmic decay time scale is generally very large because of 
the high conductivity of the matter in the core of a neutron 
star \citep{articleA}. Consequently, the field 
evolution is dominantly caused by the ambipolar diffusion of the charged particles in the core. The Hall drift does not dissipate the 
magnetic field, it only causes the change in field configuration 
inside the star. The field evolution due to ambipolar diffusion has been discussed in several recent publications \cite{2017PhRvD..96j3012G,2017MNRAS.465.3416P,2017MNRAS.471..507C,2018PhRvD..98d3007O,2020MNRAS.498.3000C}. Most of them have studied the diffusion considering the present magnetic field as perturbation. In those studies the magneto-hydrodynamic equations have been treated with different  approximations according to different physical situations suitable to neutron star core at different stages and accordingly the field evolution has been estimated. The crucial point to calculate the field evolution due to ambipolar diffusion is to have the ambipolar diffusion velocity profile inside the star core. The ambipolar velocity profile has been studied for normal and superfluid matter in non-rotating neutron star model by Passamonti et al. \citep{2017MNRAS.465.3416P}.
 
In the present work we focus on the field decay
in the core of the star caused by the ambipolar diffusion by 
evaluating the ambipolar velocity of the charged particles 
in the core without considering any approximations.
We consider the neutron star after the temperature reaches 
the values when neutrino can escape of the matter. Below this 
region of tempaerature at different temperature regions different
damping factors to ambipolar diffusion is dominant. The procedure 
discussed here is free of constraints to any such
specific temperature regime, for which some special type of 
damping of ambipolar diffusion is dominant. The way is general 
and applicable to all composition of star core at all temperature 
regimes below nutrino trapping temperature. Inside
the magnetars, due to presence of strong magnetic field the neutron
sperfluidity and protont superconductivity are quenched \citep{2014arXiv1403.2829S,
2015PhRvC..91c5805S,2016PhRvC..93a5802S}. Hence, we 
study the ambipolar diffusion velocity field and its effect 
on magnetic energy dissipation, time scale etc. generally with
all terms present, extensively for the core with normal matter.In next section 
we describe the basic theoretical calculation for solving the
ambipolar diffusion velocity in different regions of star.
We then proceed to discuss our model for which we have calculated the ambipolar velocity
in section \ref{sec:model}. We report and discuss the result
in the section \ref{sec:res}
and finally we conclude in the last section (section \ref{sec:dis}).
\section{Theoretical description}
\subsection{Particle dynamics}
The pioneer work on field decay mechanism is by Goldreich and 
Reisenegger in 1992 \citep{1992ApJ...395..250G}. For the sake  
of simplicity we consider that the matter inside the neutron star is composed of nuclear 
matter with electrons in normal phase (not in superfluid phase).
Further, we consider the star is rotating with angular speed $\Omega$ 
and the particles inside the star are in dynamical equilibrium. 
Following \citep{1992ApJ...395..250G} we assume the Newtonian 
gravitational potential $\phi$ inside the star. Then the equilibrium 
conditions of all species in the presence of electromagnetic 
field $\mathbf{E}$ and $\mathbf{B}$ are respectively,
\begin{eqnarray}
0 =m_N\mathbf{\nabla}U -\mathbf{\nabla} \mu_n + \mathbf{F}^{int}_n  %\nonumber \\
- m_N2\mathbf{\Omega}\times\mathbf{v}_n
\label{streuln} \\
0 = m_N\mathbf{\nabla}U -\mathbf{\nabla} \mu_p + e (\mathbf{E}+\mathbf{v}_p\times\mathbf{B}) +  
%\nonumber \\ 
\mathbf{F}^{int}_p - m_N2\mathbf{\Omega}\times\mathbf{v_p}
\label{streulp} \\
0 = m_e\mathbf{\nabla}U -\mathbf{\nabla} \mu_e - e (\mathbf{E}+\mathbf{v_e}\times\mathbf{B}) + \mathbf{F}^{int}_e 
%\nonumber \\ 
- m_e2\mathbf{\Omega}\times\mathbf{v}_e
\label{streule}
\end{eqnarray}
where 
\begin{equation}
	U = \mathbf{\nabla}\left(\frac{r^2\Omega^2}{2}+\phi\right),
\end{equation}
$m_i$ is the mass in the medium inside the star, $v_i$ is the velocity, $\mu_i$ is the chemical potential of the $i$-th species and $\mathbf{F}^{int}_i$
is the force on the $i$-th particle due to interaction with other 
particles. Here we denote the species neutron ($n$), proton
($p$) and electron ($e$) by the subscript $i=n,p,e$ respectively. Here we assume the masses of nucleons are same 
as $m_N$. Now multiplying the single particle equations of motion
by number density of respective species, we write the equation for
the neutrons as 
\begin{eqnarray}
	0 = m_Nn_n\mathbf{\nabla}U ~-~ n_n\mathbf{\nabla} \mu_n + \mathbf{f}^{int}_n~ 
	%\nonumber \\
	-~ m_Nn_n2\mathbf{\Omega}\times\mathbf{v}_n
\label{streulnden2}
\end{eqnarray}
and for charged particles together as
\begin{eqnarray}
	0 = (m_N+m_e)n_c\mathbf{\nabla}U ~-~ n_c(\mathbf{\nabla} \mu_p+\mathbf{\nabla} \mu_e)~~
	%\nonumber \\~&
	+~~\mathbf{f}^m + \mathbf{f}^{int}_p + \mathbf{f}^{int}_e 
	\nonumber \\
	-~ 2\mathbf{\Omega}\times n_c (m_N\mathbf{v}_p+m_e\mathbf{v}_e)
\label{strsumpeden2}
\end{eqnarray}
Here we have used the fact that $n_p=n_e=n_c$ due to charge neutrality condition where $n_i$ is number density 
of the $i$-th species. Here
\begin{equation}
    \mathbf{f}^m = \mathbf{j}\times\mathbf{B}
\end{equation}
is the magnetic force density with
\begin{equation}
    \mathbf{j} = \mathbf{j}_p + \mathbf{j}_e = n_c e (\mathbf{v}_p-\mathbf{v}_e),
    \label{ncurden}
\end{equation}
the total current density and
\begin{equation}
\mathbf{f}^{int}_i=\Sigma_jn_{ij}\mathbf{F}^{int}_{ij},
\label{coupden}
\end{equation}
with $n_{ij}$ number of collisions within unit volume between 
$i$-th and $j$-th particles, is the interaction force on $i$-th 
particle due to its interaction with others particle within 
unit volume {\it i.e.} the interaction force density for $i$-th particle.
Then interaction force density on neutrons is 
\begin{equation}
\mathbf{f}^{int}_n = n_{np}\mathbf{F}^{int}_{np}+n_{ne}\mathbf{F}^{int}_{ne}
\label{cfricden}
\end{equation}
and on charged particles
\begin{eqnarray}
	\mathbf{f}^{int}_p+\mathbf{f}^{int}_e & =\left(n_{pn}\mathbf{F}^{int}_{pn}+n_{pe}\mathbf{F}^{int}_{pe}\right) 
	%\nonumber \\
	+ \left(n_{pe}\mathbf{F}^{int}_{ep}+n_{en}\mathbf{F}^{int}_{en}\right) 
	\nonumber \\ &	
	= n_{pn}\mathbf{F}^{int}_{pn} + n_{en}\mathbf{F}^{int}_{en} 
\end{eqnarray}
Hence, we have
\begin{equation}
	\mathbf{f}^{int}_p+\mathbf{f}^{int}_e = -\mathbf{f}^{int}_n = -\mathbf{f}^{int}~{\rm (say)}
\end{equation}
Then dividing the eq. (\ref{streulnden2}) by neutron mass density 
$\rho_n$ and eq. (\ref{strsumpeden2}) by proton mass density
$\rho_p$ we have
\begin{eqnarray}
0=\mathbf{\nabla}U ~-~ \mathbf{\nabla}\left(\frac{\mu_n}{m_N}\right)+\frac{\mathbf{f}^{int}}{\rho_n} ~ 
\nonumber \\
-~ 2\mathbf{\Omega}\times\mathbf{v}_n
   \label{strnacc}
\\
    0 = \mathbf{\nabla}U ~-~\mathbf{\nabla}\left(\frac{\mu_p+\mu_e}{m_N}\right)+\frac{\mathbf{f}^m}{\rho_p}-\frac{\mathbf{f}^{int}}{\rho_p} ~
    \nonumber \\
    -~ 2\mathbf{\Omega}\times \mathbf{v}_p
    \label{strsumpeacc}
\end{eqnarray}
as $m_e\ll m_N$.

Then from the difference of these equations we get the combined 
equation of motion as
\begin{equation}
	0 =x_p\mathbf{\nabla}\beta ~+~ \frac{\mathbf{f}^{int}}{\rho_n} ~+~ x_p2\mathbf{\Omega}\times \mathbf{V}_p  ~-~ \frac{\mathbf{f}^m}{\rho} \label {strsumpedeffnacc}
\end{equation}
where $\beta=\left(\mu_p+\mu_e-\mu_n\right)/m_N$ and $\mathbf{V}_p=\mathbf{v}_p-\mathbf{v}_n$ is
the proton velocity in neutron rest frame. Then in absence of rotation 
the magnetic force is balanced by force due to (1) Transfusion
and (2) Interaction.
 
\subsubsection{Transfusion}

When due to motion of charged particles the chemical equilibrium 
is disturbed, the weak reaction 
\begin{equation}
    n~\longleftrightarrow~p~+~e
    \nonumber
\end{equation}
tries to restore the chemical equilibrium. In this case, the number 
of particles generated or destroyed per unit time is given by 
reaction rate $\Gamma=\lambda\beta$. The charged particles follow
the equation of continuity
\begin{equation}
    \frac{\partial n_i}{\partial t}+\mathbf{\nabla}\cdot \left(n_i \mathbf{v}_i\right) = -\lambda\beta
\end{equation}
and as $n_p=n_e=n_c$ we have \citep{1992ApJ...395..250G}
\begin{equation}
    \mathbf{\nabla}\cdot \left(n_c \mathbf{w}\right)
= -\lambda\beta 
\label{transfusion}
\end{equation}
where
\begin{equation}
    \mathbf{w} = \frac{\mathbf{v}_p+\mathbf{v}_e}{2}
    \label{avchvel}
\end{equation}

\subsubsection{Interaction}

In absence of superfluidity and superconductivity, the interaction
between species are basically mechanical collision between them
and that can be expressed as proportional to their relative 
velocity as
\begin{equation}
    \mathbf{F}^{col}_{ij}=-{\cal D}_{ij}(\mathbf{v}_i-\mathbf{v}_j)
\end{equation}
where ${\cal D}_{ij}$ is the collision coefficient for collision
between $i$-th and $j$-th species.

Hence,
\begin{eqnarray}
\mathbf{f}^{int} = \mathbf{f}^{int}_n = n_{np}{\cal D}_{np}\mathbf{V}_p ~
    + n_{ne}{\cal D}_{ne}\mathbf{V}_e 
    \nonumber \\
    = n_{np} m_n \tau_{np}\mathbf{V}_p ~
    + n_{ne} m_n\tau_{ne}\mathbf{V}_e 
    \nonumber \\
    = D_{np}\mathbf{V}_p + D_{ne}\mathbf{V}_e 
	\nonumber \\
	= (D_{np} + D_{ne})\mathbf{V}
    \end{eqnarray}
where, $D_{ij}=n_{ij}m_n\tau_{ij}$ with $\tau_{ij}$ the collision rate between $i$-th and $j$-th species, and $\mathbf{V}_i$ is velocity 
of $i$-th species in rest frame of neutron $\mathbf{v}_i-\mathbf{v}_n$.
Here we define the ambipolar velocity 
\begin{equation}
	\mathbf{V} = \frac{D_{np}\mathbf{V}_p + D_{ne}\mathbf{V}_e}{D_{np} + D_{ne}}.
\end{equation}

\subsection{Field evolution}

Now in a plasma due to ambipolar diffusion 
\citep{1992ApJ...395..250G} the magnetic field evolution is given by
\begin{equation}
	\frac{\partial \mathbf{B}}{\partial t} = \mathbf{\nabla}
	\times \left(\mathbf{V} \times \mathbf{B}\right)
\end{equation} and the decay in magnetic energy is given by
\begin{equation}
	\frac{\partial E_B}{\partial t} = -\frac{1}{4\pi } \int d^3 x \mathbf{V} 
	\cdot \mathbf{f}^m
	\label{magnetic}
\end{equation}
where $\mathbf{V}$ is the ambipolar diffusion velocity.

\section{Model}
\label{sec:model}
We consider a star composed of normal nucleons and electrons. 
We assume the magnetic field of the star is purely poloidal 
as $\mathbf{B}=B_\zeta\hat{\zeta} + B_z \hat{z}$ in cylindrical
coordinate system with $\hat{\zeta}$ as the unit vector in the 
direction of cylindrical radius. We choose the rotation axis 
as the positive $z$ axis and the system has axial symmetry. 
With this choice of configuration we solve the dynamical equilibrium 
equation for ambipolar velocity $\mathbf{V}$ as
\begin{eqnarray}
	0 = x_p\mathbf{\nabla}\beta ~+~ \frac{D}{\rho_n} \mathbf{V}
~+~ x_p2\mathbf{\Omega}\times (\mathbf{V} + A \mathbf{j}) 
	%~\nonumber \\
	-~ \frac{\mathbf{f}^m}{\rho}.
	\label{eqmot1}
\end{eqnarray}
where $D = D_{np} + D_{ne}$. \\
To arrive at the above equation we express proton and electron
velocities ($\mathbf{v}_p,\mathbf{v}_e$) in the eq. (\ref{avchvel}) 
in terms of ambipolar velocity $\mathbf{V}$ and current density 
$\mathbf{j}$ with
\begin{equation}
	A = \frac{D_{ne}}{n_{ce}\left( D_{np} +D_{ne}\right)}.
\end{equation}

Expressing the velocities of species in terms of ambipolar velocity 
$V$ we can express the chemical in-equilibrium as
\begin{equation}
    \mathbf{\nabla} \beta = -\frac{1}{\lambda}\mathbf{\nabla} \left\{\mathbf{\nabla} \cdot \left(n_c \mathbf{V}\right)\right\}
\end{equation}
from eq. (\ref{transfusion}). Substituting this in eq. (\ref{eqmot1}), 
we get
\begin{eqnarray}
    	0 = -\frac{x_p}{\lambda}\mathbf{\nabla} \left\{\mathbf{\nabla} \cdot \left(n_c \mathbf{V}\right)\right\} ~+~ \frac{D}{\rho_n} \mathbf{V}
~
%\nonumber \\
+~ x_p2\mathbf{\Omega}\times (\mathbf{V} + A \mathbf{j}) 
	~-~ \frac{\mathbf{f}^m}{\rho}.
	\label{amb}
\end{eqnarray}
Eq. (\ref{amb}) is an inhomogeneous second order differential 
equation in $\mathbf{V}$ for a rotating neutron star with a 
magnetic field $\mathbf{B}$. We try to solve this equation with boundary conditions for ambipolar velocity and its gradient to vanish at the centre of the star. To obtain its solution, we assume 
the present magnetic field configuration to be completely poloidal 
which can be obtained from the magnetic vector potential 
\begin{equation}
    \mathbf{A} = \frac{1}{2}\zeta B\sin\theta \hat{\phi}
\end{equation}
where the magnitude of the magnetic field $B$ varies inside 
the star with the density profile \citep{Bandyopadhyay:1997kh}, 
parametrised as:
\begin{equation}
    B\left(\frac{n_b}{n_0}\right) = B_s + B_c\left[1-\beta\left(\frac{n_b}{n_0}\right)^\gamma\right],
\end{equation}
where $n_0$ is the nucleon saturation density, $n_b$ the baryonic number
density and $B_s$ and $B_c$ 
are the magnitude of the magnetic field at the surface and core of the star 
respectively. 

\begin{figure}
    \centering
    \includegraphics[width=8cm]{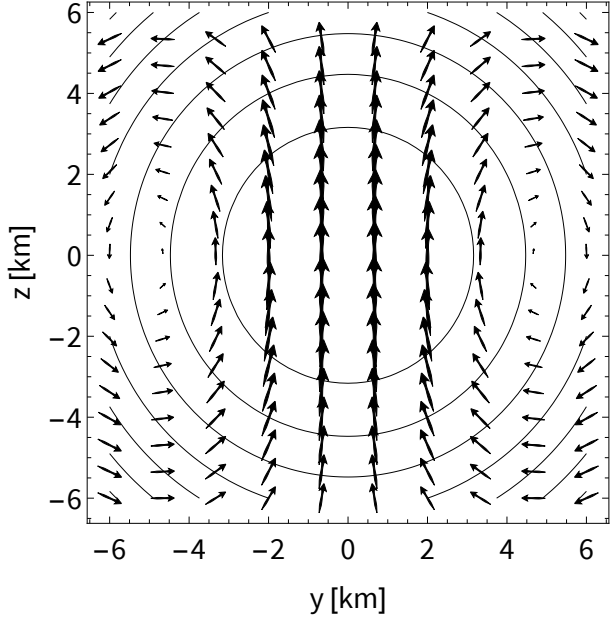}
    \caption{ Density dependent poloidal magnetic field inside the  neutron star}
    \label{fig:Magnetic Field}
\end{figure}

To obtain the solution of ambipolar velocity from eq. (\ref{amb}) 
further we consider a model star with matter inside the star 
composed of only nucleons with electrons and equation of state 
of the matter constructed within relativistic mean field model 
with GM1 parametrisation \citep{PhysRevLett.67.2414}. To be specific we consider 
a star with mass $2.217$ M$_\odot$ and radius $10.614$ km. With 
the magnetic field configuration described above the field profile 
inside the star is shown in figure \ref{fig:Magnetic Field}. To 
get this profile the value of $\beta$ and $\gamma$ are set to 
$0.01$ and $2.0$ respectively.

We now solve for the ambipolar diffusion velocity ($\mathbf{V}$) 
in the core of the star taking into consideration the above 
magnetic field. The reaction rate and the collision coefficients are assumed constant throughout the star. To evaluate these constants, we consider a typical matter density of $2.2n_0$. The corresponding charged 
particle densities, neutron density and their fractions are 
shown in table \ref{tab:den}. However, taking the actual values of these quantities at different points of the star, the solution for ambipolar velocity profile can also be obtained by a more general numerical method.

\begin{table*}[]
    \centering
     \caption{Used matter density}
    \begin{tabular}{|c|c|c|c|c|c|}
    \hline
     $x_p$ & $x_n$ & $\rho_n$ & $\rho$ & $n_c$  & $n_B$  \\
 &  & g cm$^{-3}$  &  g cm$^{-3}$ & fm$^{-3}$ &  fm$^{-3}$ \\
\hline
$0.1346$ & $0.8692$  & $4.435\times10^{14}$ & $10.2204\times10^{14}$ & $0.044$ & $0.3366$ \\
 \hline
    \end{tabular}
    \label{tab:den}
\end{table*} 

For $x_p$ $\geq$ $0.11$, $\mathbf{\beta}$-reactions would be driven by direct Urca(dUrca) processes \citep{PhysRevLett.66.2701}. For this process, the reaction rate is given by \citep{PhysRevD.45.4708}
\begin{eqnarray}
    \lambda = 3.5 \times 10^{36}\frac{m_n^*}{m_n}\frac{m_p^*}{m_p}T_8^4\left(\frac{\rho}{\rho_{nuc}}\right)^\frac{1}{3} 
    %\nonumber \\
    ~{\rm erg}^{-1}{\rm cm}^{-3}{\rm s}^{-1}
\end{eqnarray}

The collision coefficients are obtained from \citep{articleA}, who calculated the collision frequencies and proved that the neutron-proton scattering is much more frequent than the neutron-electron scattering

\begin{equation}
    \tau_{np} = 6.6 \times 10^{16}T_8^2\rho_{14}^{-\frac{1}{3}} ~{\rm s}^{-1}
\end{equation}
Here one should note that as an example of the solution procedure we have taken here a typical density of matter and other necessary quantities corresponding to that density. However, in actual calculation every input parameter required to solve should be taken as function of density. However, the the nature of result does not differ much except nominal quantitative corrections. 

The angular velocity is calculated using the time-period of rotation of a magnetar which is obtained from Olausen and Kaspi \citep{Olausen_2014}. We consider the value:

\begin{equation}
    \Omega = 0.628 ~\rm{rad/s}
\end{equation}

\section{Results}\label{sec:res}

\begin{figure*}
    \centering
    \includegraphics[width=14cm,height=5.0cm]{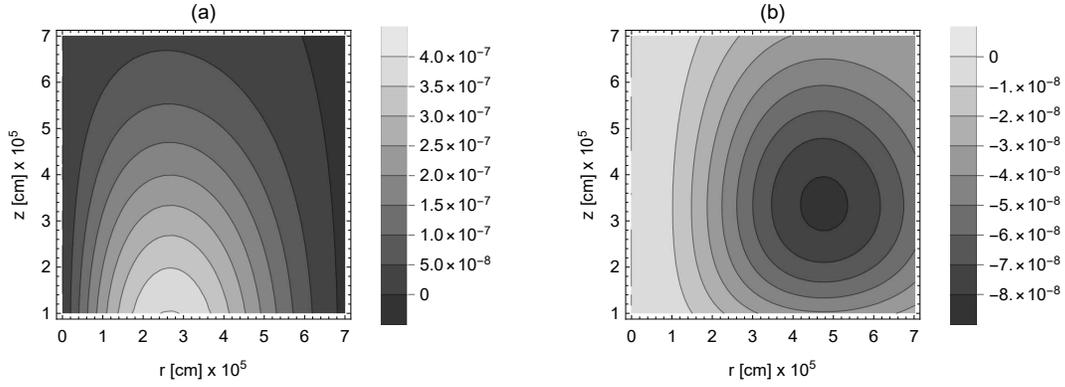}
     \caption{The radial and axial components of ambipolar diffusion velocity  in the upper hemisphere of the neutron star core (z $>0$). (a): the radial component of ambipolar diffusion velocity, (b): The axial component of ambipolar diffusion velocity}
     \label{fig:Vr}
\end{figure*}

\begin{figure*}
    \centering
    \includegraphics[width=14cm,height=5cm]{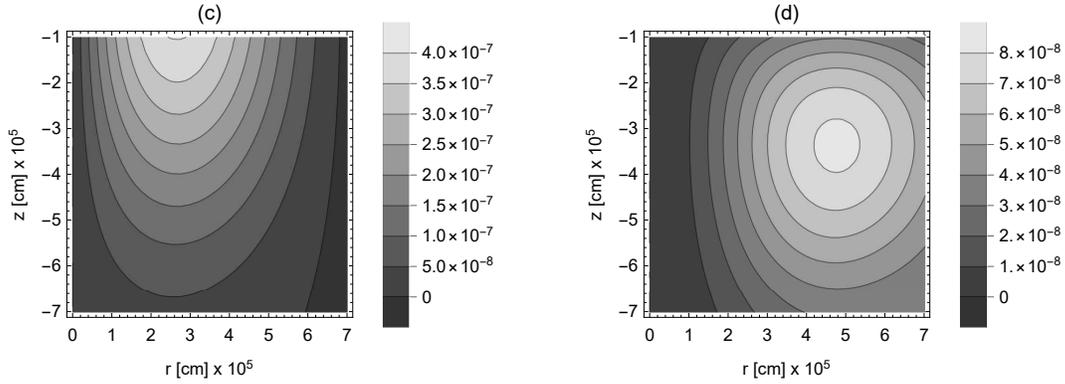}
     \caption{Same as in figure\ref{fig:Vr} in the lower hemisphere of the neutron star core (z $<0$).}
     \label{fig:Vz}
\end{figure*}

\begin{figure}
    \centering
    \includegraphics[width=8cm]{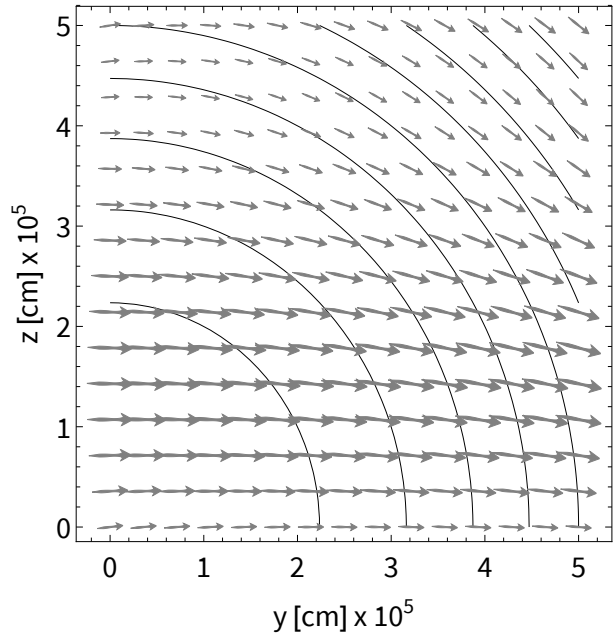}
    \caption{The ambipolar diffusion velocity vector in the core of the neutron star. The circular arcs represents the region of different radius inside the neutron star core . The arrow represents the direction of the velocity vector.}        
    \label{fig:V}
\end{figure}

With the above described model of star with matter equation of state within relativistic mean field model with GM1 parametrization \citep{PhysRevLett.67.2414} and the density dependent purely poloidal magnetic field, a numerical solution of Eq. (\ref{amb}) is carried out to obtain the profile of the ambipolar velocity $\mathbf{V}$. We carry out our calculations for a particular temperature $T=8\times10^8$ K which falls within the range of typical temperature inside neutron star core in the early age of the stars. The radial and axial components of the velocity vector has been shown in figure. \ref{fig:Vr} and figure. \ref{fig:Vz} for upper and lower hemisphere respectively. The azimuthal component has the same profile (a consequence of the poloidal nature of magnetic field) as the radial component, its magnitude being negligible. The total velocity profile in a quadrant part of the neutron star core is shown in figure. \ref{fig:V}.

The time-scale of magnetic field evolution as caused by ambipolar diffusion can be estimated as
\begin{equation}
    t = \frac{L}{\langle V \rangle}
\end{equation}
where $L$ is the characteristic length over which magnetic field varies and $\langle V \rangle$ is the volume average of the magnitude of the diffusion velocity. For a characteristic length of 8 kms, we get a timescale of $\approx$ $9.42 \times 10^4$ years. The energy dissipation rate due to ambipolar diffusion can be found by substituting $\mathbf{V}$ in eq. (\ref{magnetic}). For the velocity vector shown in figure. \ref{fig:V}, we obtain the magnetic energy dissipation rate 
\begin{equation}
    \frac{dE_B}{dt} = - 1.22 \times 10^{36} ~\rm{ergs/s} .   
\end{equation}
The negative sign indicates the decrease in magnetic energy.

\section{Discussion}\label{sec:dis}

In the present work, we outline the general method to obtain the magnetic energy decay rate due to ambipolar diffusion in the core of a compact star. In some of the existing literature \citep{2017PhRvD..96j3012G,2017MNRAS.471..507C,2018PhRvD..98d3007O,2020MNRAS.498.3000C}, the field evolution has been computed using perturbation method and discussed the importance of ambipolar diffusion process in the magnetic field evolution. With this approach the background baryon velocity completely depends on boundary condition of magnetic field profile which in turn depends on rate of field evolution and that is unknown a priory. Moreover, the ambipolar velocity has been determined as an estimate from the baryon background velocity \citep{2018PhRvD..98d3007O,2020MNRAS.498.3000C}. They consider the chemical potential variation inside the star as it is in equilibrium, with no variation of the chemical potential due to drift of charged particles \citep{2017PhRvD..96j3012G,2017MNRAS.471..507C} arguing that chemical imbalance is not caused by the $\beta$-equilibrium reactions, rather controlled by magnetic field profile only, which is quite debatable.  Moreover, the approximation of equality of all particle velocity has been considered which is not required to consider in present calculation. On the other hand, \citep{2020MNRAS.498.3000C} considers the low temperature limit only in which the transfusion can be neglected. In addition, the true nature of the matter considering the presence of strong interaction has not been taken into account \citep{2020MNRAS.498.3000C}.

The magnetic energy dissipation due to ambipolar diffusion is 
related to ambipolar velocity of charged particles. In this 
context we have constructed the equations from simple microscopic 
idea of particle dynamics in a normal matter composed of proton-electron 
plasma in the background of neutrons and in the presence of 
external magnetic field. Equations contain terms which are 
sensitive to temperature and density. To get a complete picture 
in the star core interior for different temperature regime, 
it is necessary to solve the equations for a wide range to 
temperature and matter density with different possibilities 
of matter properties and composition which is to be addressed 
in near future. To get the general idea of solution procedure, 
at some specific matter density compatible to the core of the 
compact stars,  we solve the dynamical equilibrium equation 
to get ambipolar diffusion velocity profile inside a comparatively 
young neutron star core with typical temperature of $\sim8\times10^8$ 
K. As temperature is high inside the core,it is quite natural 
for the matter in the core to be in non-superfluid phase.
Moreover, in presence of ultra strong magnetic
field inside the magentars, the neutron superfluidity and
proton superconductivity are quenched \citep{2014arXiv1403.2829S,
2015PhRvC..91c5805S,2016PhRvC..93a5802S}. 

From 
this point of view we consider the normal matter in this current 
calculation. From the profile of the ambipolar velocity we 
can get the idea of magnetic energy dissipation due to ambipolar 
diffusion, the typical time scale of magnetic field decay and 
also the amount of heating due to field decay. We have shown 
that with a realistic model, the energy dissipation rate is 
$\sim 10^{36}$ ergs $s^{-1}$ as expected \citep{1995MNRAS.275..255T}. 
We obtain a value of magnetic energy dissipation as $\sim 3.64 
\times 10^{48}$ ergs during the timescale of evolution which 
reproduce the order of total magnetic energy budget as mentioned 
in literature by \citep{review}. The time scale of magnetic 
field decay is $\sim 10^4-10^5$ yrs, which is compatible with 
the observed lifetime of soft gamma repeaters activity.  One 
should note that our model considers the core to be composed
of normal (non-superfluid) matter which is very possible scenario 
if the star posses very strong magnetic field like magnetars 
\citep{2014arXiv1403.2829S, 2015PhRvC..91c5805S,2016PhRvC..93a5802S}.
We take an approximation by considering a purely poloidal magnetic 
field. An additional toroidal magnetic field will not only 
change the profile of azimuthal component of $\mathbf{V}$, 
but will also increase the magnitude of the radial and the 
axial components. The increased magnitude of $\mathbf{V}$ shall 
decrease the time-scale of evolution possibly to $10^3-10^4$ yrs. 

In this work we have not taken any approximation considering the temperature regime. We have discussed the result for a particular temperature but the solution procedure is not specific to that temperature. The procedure is general which can be employed for any temperature. Ambipolar velocity profile will be different for different temperature region because of temperature dependence of various parameters involved in the differential equation of ambipolar velocity. In this respect, this work shows a general procedure to solve the ambipolar diffusion problem to get an exact picture of the field decay rate and heating due to field decay with different external realistic conditions relevant at different thermal and temporal stage of the stars. The work shows a general procedure to get the ambipolar velocities without any approximation or assumptions of special case and a solution with a typical model of star containing normal matter as an example. 

Natural extension of the work is to get field decay in all possible scenario associated with the neutron star observations to date.  The detail calculation of decay rate at different stages of neutron star during its thermal evolution considering different phases of matter inside the core of the star requires extensive study of the subject keeping in mind the different external conditions. It includes the cases, superfluidity of neutrons, superconductivity of protons and in some cases of magnetars with the quenching of superconductivity. All these conditions together will determine the thermal and field evolution and hence the persistent luminosity of some class of neutron stars. Finally we should emphasize that the obtained numerical values of different physical quantities are not final because we have not considered real system as a whole, though the procedure is general and to get the real picture we need extensive computational work considering the whole system together with proper input of temporal evolution.

\section*{Acknowledgements}

The authors acknowledge the funding support from Science and Engineering Research Board, Department of Science and Technology, Government of India through Project No. EMR/2016/006577.  

%%%%%%%%%%%%%%%%%%%%%%%%%%%%%%%%%%%%%%%%%%%%%%%%%%
%\section*{References}
%\bibliographystyle{unsrt}
\bibliography{ambipolar}

\end{document}